\documentclass{aa}

\usepackage{amssymb}
\usepackage{amsmath}
\usepackage{graphicx}
\usepackage{txfonts}

\usepackage{textcomp}
\bibpunct{(}{)}{;}{a}{}{,}

\begin{document}

\title{A search for \textit{Fermi} bursts associated to supernovae and their frequency of occurrence}
\titlerunning{\textit{Fermi} bursts and supernovae}
\authorrunning{Kovacevic, Izzo, Wang et al.}
\author{M. Kovacevic$^{1,2}$,  L. Izzo$^{1,3}$, Y. Wang$^{1}$,  M. Muccino$^{1,3}$, M. Della Valle,$^{3,4}$,  L. Amati$^{5}$, C. Barbarino$^{1,4}$\\
M. Enderli$^{1,2}$, G. B. Pisani$^{1,2}$,  L. Li$^{6,7,8}$}
\offprints{\email{milosh.kovacevic@gmail.com}\\ \email{luca.izzo@gmail.com,wangyu@me.com}}
\institute{$^{1}$Dip. di Fisica and ICRA, Sapienza Universit\`a di Roma, Piazzale Aldo Moro 5, I-00185 Rome, Italy.\\
$^{2}$Universit\'e de Nice Sophia Antipolis, CEDEX 2, Grand Ch\^{a}teau Parc Valrose, Nice, France.\\
$^{3}$ICRANet-Pescara, Piazza della Repubblica 10, I-65122 Pescara, Italy.\\
$^{4}$INAF-Napoli, Osservatorio Astronomico di Capodimonte, Salita Moiariello 16, I-80131 Napoli, Italy.\\ 
$^{5}$INAF, Istituto di Astrofisica Spaziale e Fisica Cosmica, Bologna, Via Gobetti 101, I-40129 Bologna, Italy\\
$^{6}$Department of Physics, KTH Royal Institute of Technology, AlbaNova University Center, SE-106 91 Stockholm, Sweden.\\
$^{7}$The Oskar Klein Centre for Cosmoparticle Physics, AlbaNova, SE-106 91 Stockholm, Sweden.\\
$^{8}$Department of Physics Stockholm University, AlbaNova, SE-106 91 Stockholm, Sweden.}
\date{}

\abstract
{Observations suggest that the major fraction of long duration gamma-ray bursts (GRBs) are connected with broad-lines supernovae Ib/c, (SNe-Ibc). The presence of GRB-SNe is revealed by rebrightenings emerging from the optical GRB afterglow $10$--$15$ days, in the rest-frame of the source, after the prompt GRB emission.}
{\textit{Fermi}-GBM has a field of view (FoV) which is about 6.5 times larger than the FoV of \textit{Swift}, therefore we expect that a number of GRB-SN connections have been missed due to lack of optical and X-ray instruments on board of \textit{Fermi}, which are essential  to reveal SNe associated with GRBs. This fact has motivated our search in the \textit{Fermi} catalogue for possible GRB-SN events.}  
{The search for possible GRB-SN associations follows two requirements: (1) SN should fall inside the \textit{Fermi}-GBM error box of the considered long GRB, and (2) this GRB should occur within $20$ days before the SN event.}
{We have found $5$ cases, within $z<0.2$  fulfilling the above reported requirements. One of them, GRB 130702A-SN 2013dx,  was already known as GRB-SN association. We have analyzed the remaining $4$ cases and we concluded that three of them are, very likely, just random coincidences, due to the \textit{Fermi}-GBM large error box associated with each GRB detection. We found one GRB possibly associated with a SN 1998bw-like, GRB 120121B/SN 2012ba.}
{The very low redshift of GRB 120121B/SN 2012ba ($z = 0.017$) implies a low isotropic energy of this burst ($E_{iso} = 1.39 \times 10^{48}$) erg. We then compute the rate of \textit{Fermi} low-luminosity GRBs connected with SNe to be $\rho_{0,b} \leq 770\ $Gpc$^{-3}$yr$^{-1}$. We estimate that \textit{Fermi}-GBM could detect $1$--$4$ GRBs-SNe within $z \leq 0.2$ in the next 4 years.}


\maketitle

\section{Introduction}

Gamma-ray bursts (GRBs) are the most powerful stellar explosions in the universe {\citep[see][for a review]{Piran2005, GehrelsMeszaros2012,Zhang2014}, with a total isotropic energy release of $E_{iso} = 10^{48-54}$ erg. Their origin is associated to the final collapse of a very massive star or to the merging of two compact objects. This first taxonomy was inferred from the existence of two observed classes for GRBs, based on their $T_{90}$ duration \citep{Klebesadel1982,Dezalay1992,Koveliotou1993,Tavani1998}: GRBs with $T_{90}<2$ s are named short GRBs; otherwise they are named long GRBs. All GRBs associated with supernovae (SNe) have been confirmed to be long bursts, although the opposite might not be true \citep{DellaValle2006b,Fynbo2006,Gal-Yam2006}.  Observations carried out in the last decade suggest that long GRBs are associated with SNe Ib/c, which are believed to originate from the collapse of single very massive stars \citep{Heger2003} or from moderate mass Wolf-Rayet stars in interacting binaries \citep{Smartt2009}. To date, $35$ GRB-SN associations have been confirmed on spectroscopic and/or photometric grounds, see Table \ref{tab:no0}.  The SN lightcurve peaks at $10$--$15$ days after the GRB trigger (in the source rest-frame) powered by the radioactive decay of $^{56}$\ion{Ni}, and whose half-life time is about $6$ days \citep{Arnett1996}. 

\begin{table*}
\centering
\tiny{
\begin{tabular}{lcccccc}
\hline\hline
GRB & $E_{iso}$ & Discovered by &  $z$   & SN & SN & Refs.\\
 & $(erg)$ &  &    & identification & name & \\
\hline
970228 & 1.86 $\times$ 10$^{52}$ & BATSE/SAX & 0.695 & bump & & \citep{Reichart1997} \\
980326 & 5.60 $\times$ 10$^{51}$ & BATSE/SAX & 1(?) & bump & &  \citep{Bloom1999} \\
980425 & 6.38 $\times$ 10$^{47}$& BATSE & 0.0085 & spec. & SN1998bw & \citep{Galama1998} \\
990712 & 7.80 $\times$ 10$^{51}$& SAX & 0.434 & bump & & \citep{Frontera2009,Zeh2004} \\
991208 & 2.59 $\times$ 10$^{53}$& SAX & 0.706 & bump & & \citep{Frontera2009,Zeh2004} \\
000911 & 7.80 $\times$ 10$^{53}$& Konus-WIND & 1.058 & bump & & \citep{Lazzati2001,Hurley2000} \\
010921 & 1.10 $\times$ 10$^{52}$& HETE & 0.45 & bump & & \citep{Zeh2004}\\
011121 & 9.90 $\times$ 10$^{52}$& Ulysses & 0.36 & bump & SN 2001ke & \citep{Bloom2002,Hurley2001,Greiner2003}\\
020305 & 0.7-4.6 $\times$ 10$^{51}$& Ulysses & 0.2-0.5 & bump & & \citep{Gorosabel2005,Hurley2002}\\
020405 & 1.28 $\times$ 10$^{53}$& Ulysses & 0.695 & bump & & \citep{Masetti2003,Hurley2002b}\\
020410 & 2.20 $\times$ 10$^{52}$& Konus-WIND & $\sim$ 0.5 & bump & & \citep{Nicastro2004,Levan2005}\\
021211 & 1.30 $\times$ 10$^{52}$& HETE & 1.006 & spec. & SN 2002lt & \citep{dellavalle2003,Vreeswijk2003,Crew2002}\\
030329 & 1.70 $\times$ 10$^{52}$& Konus-WIND & 0.168 & spec. & SN 2003dh & \citep{Golenetskii2003,Kawabata2003,Stanek2003}\\
030723 & < 1.60 $\times$ 10$^{53}$ & HETE & < 1 & bump & & \citep{Fynbo2003}\\
031203 & 9.99 $\times$ 10$^{49}$& INTEGRAL & 0.105 & spec. & SN 2003lw & \citep{Soderberg2003,Tagliaferri2004}\\
040924 & 1.10 $\times$ 10$^{52}$ & HETE & 0.86 & bump & & \citep{Fenimore2004,Soderberg2006b}\\
041006 & 3.50 $\times$ 10$^{52}$& HETE & 0.716 & bump & & \citep{Galassi2004,Bikmaev2004,Soderberg2006b}\\
050525A & 3.39 $\times$ 10$^{52}$& Konus-WIND & 0.606 & spec. & SN 2005nc & \citep{DellaValle2006}\\
060218 & 1.66 $\times$ 10$^{49}$& Swift & 0.033 & spec. & SN 2006aj & \citep{Campana2006,Soderberg2006}\\
060729 & 1.60 $\times$ 10$^{52}$& Swift & 0.54 & bump & & \citep{Cano2011,Parsons2006}\\
070419 & 7.90 $\times$ 10$^{51}$& Swift & 0.97 & bump & & \citep{Hill2007}\\
080319B & 1.30 $\times$ 10$^{54}$& Swift & 0.937 & bump & & \citep{Perley2008,Kann2008,Cummings2008}\\
081007 & 2.50 $\times$ 10$^{51}$& Swift & 0.5295 & bump& SN2008hw & \citep{Soderberg2008,Markwardt2008}\\
090618 & 2.90 $\times$ 10$^{53}$& Fermi-GBM & 0.54 & bump & & \citep{Izzo2012,Cano2011,GCN9535}\\ 
091127 & 1.60 $\times$ 10$^{52}$& Fermi-GBM & 0.49 & bump & SN 2009nz & \citep{Cobb2010,Wilson-Hodge2009}\\
100316D & 9.81 $\times$ 10$^{48}$& Swift & 0.059 & spec. & SN 2010bh & \citep{Bufano2012,Chornock2010,Sakamoto2010}\\
101219B & 4.39 $\times$ 10$^{51}$& Fermi-GBM & 0.55 & spec. & SN 2010ma & \citep{Sparre2011,vanderhorst2010}\\
111228A & 7.52 $\times$ 10$^{52}$& Fermi-GBM & 0.714 & bump & & \citep{D'Avanzo2012,Briggs2011}\\
120422A & 1.28 $\times$ 10$^{51}$& Swift & 0.283 & spec. & SN 2012bz & \citep{Melandri2012,Barthelmy2012}\\
120714B & 4.51 $\times$ 10$^{51}$& Swift & 0.3984 & spec. & SN 2012eb & \citep{Cummings2012,Klose2012}\\
120729A & 2.30 $\times$ 10$^{52}$& Swift & 0.80 & bump & & \citep{Cano2014,Ukwatta2012}\\
130215A & 3.10 $\times$ 10$^{52}$& Fermi-GBM & 0.597 & spec. & SN 2013ez & \citep{deugarte2013,Younes2013}\\
130427A & 9.57 $\times$ 10$^{53}$& Fermi-GBM & 0.3399 & spec. & SN 2013cq & \citep{Melandri2014,Xu2013,vonkienlin2013}\\
130702A & 7.80 $\times$ 10$^{50}$& Fermi-GBM & 0.145 & spec. & SN 2013dx & \citep{Cenko2013,Collazzi2013,Singer2013}\\
130831A & 4.56 $\times$ 10$^{51}$& Konus-WIND & 0.4791 & spec. & SN 2013fu & \citep{Klose2013,Golenetskii2013}\\
\hline
\end{tabular}}
\caption{The sample of the 35 confirmed GRB-SN connections updated to the 31 May 2014.}
\label{tab:no0}
\end{table*}

With the launch of satellites dedicated to GRBs studies, as the \textit{Swift} mission \citep{Gehrels2009}, and the \textit{Fermi} spacecraft \citep{Meegan2009}, we made a step forward towards the understanding of GRB emission in the energy range between 0.3 keV up to $\sim$ 10 MeV. On the other hand, the Burst Alert Telescope \citep[BAT,][]{Barthelmy2005} on-board \textit{Swift}, is able to observe only a fraction of the sky which is 6.5 times smaller than the one covered by the \textit{Fermi}-Gamma Ray Burst Monitors (GBM) detectors \citep{Meegan2009}. This fact implies that there could exist long bursts, possibly connected with SNe, which have been detected by \textit{Fermi}-GBM without soft X-rays and optical follow up, which are essential to reveal the presence of a SN in the GRB afterglow (e.g. Mangano et al. 2007). We can make a first order estimate of the expected number of \textit{Fermi} long bursts connected with SNe as it follows. If we restrict, for reason of completeness, our analysis to GRB-SNe within $z \leq 0.2$, we have that \textit{Swift}-BAT has detected, to date,  two such events \citep[GRB 060218,][]{Campana2006} and \citep[GRB 100316D,][]{Starling2011}. Therefore, \textit{Fermi}-GBM should have discovered $2^{+2.6}_{-1.3} \times \rho_{GBM}/ \rho_{BAT} \times 0.6 = 2$--$11 \times 0.6 \sim 1$--$7$ GRB-SNe within $z\leq0.2$.  The ratio $\rho_{GBM}/ \rho_{BAT}$ ($\rho_{GBM} = 238$ GRBs yr$^{-1}$, \citealp{vonKienlin2014}, $\rho_{BAT} = 95$ GRBs yr$^{-1}$, \citealp{Sakamoto2011}) takes into account the different sky coverage of both detectors and their different sensitivities \citep{Band2003}, while the scale factor 0.6 accounts for the fact that \textit{Fermi} is monitoring the sky since about $6$ years, while \textit{Swift} since $10$ years. The attached 1$\sigma$ poissonian uncertainty at the rate of 2 GRB-SNe yr$^{-1}$, within $z \leq 0.2$, has been derived from \citet{Gehrels1986}. 

We present in Section 2 the adopted strategy that we have used to identify GRB-SN candidates. In Section 3 we discuss the 11 GRB-SN coincidences pinpointed by our code. In Section 4 we discuss our results and in Section 5 we present our conclusions.

 \section{Methodology and statistical analysis}

Our code compares the positions of the Harvard catalog of SNe\footnote{http://www.cbat.eps.harvard.edu/lists/Supernovae.html}, and the Asiago SN catalog \citep{Asiagocatalog}, with the positions of $1147$ long GRBs detected up to May 31$^{th}$ 2014, and reported in the \textit{Fermi}-GBM catalog\footnote{http://heasarc.gsfc.nasa.gov/db-perl/W3Browse} with the attached error boxes. Subsequently, we consider only  GRBs  which were detected within $\Delta t$ days before the occurrence of the SN. The exact value of $\Delta t$ days was computed after taking into account several factors: the rise time of the SN (typically 10-15 days), the assumption that GRB and SN  are simultaneous \citep{Campana2006}, and also the possibility that the SN was discovered after its maximum light. To discern physical GRB-SN associations by random spatial and temporal GRB-SN coincidences due to the large error box associated with GRB detections or uncertainties on the epoch of SN maximum,  we have computed the statistical significance of GRB-SN associations also for  SN types for which we know ``a priori'' that they are not associated with GRBs, like SNe-Ia and type II (see \citealp{Valenti2005}). In Table \ref{tab:no3} we list in the first row the assumed $\Delta t$ (in days) after the GRB trigger.  In the following rows we list the cumulative number of possible associations, within $\Delta t$, for each type of SN, respectively $N_{Ib/c}(\Delta t)$, $N_{Ia}(\Delta t)$, $N_{IIp}(\Delta t)$ and $N_{IIn}(\Delta t)$, and in the last one for all types, $N_{tot}(\Delta t)$. In the last column is also shown the percentage of the total number of each SN type over the total sample.

\begin{table*}\centering
\tiny{
\caption{The cumulative number of each SN type associated within the error radius of \textit{Fermi}-GRBs at different time intervals after the trigger time. In the first row the considered time intervals (in days) are listed. In the following rows the number of possible associations for each type of SN, respectively Ib/c, Ia, IIp and IIn, and the total number of SNe, for each considered time interval, are listed. In the last column the percentage $r_x$ of the total number of each SN type over the total sample is shown.}
\begin{tabular}{c|ccccccccccccccccc|c}     
\hline\hline
$\Delta t$ (days) & 10 & 20 & 30 & 40 & 50 & 60 & 70 & 80 & 90 & 100 & 110 & 120 & 150 & 200 & 300 & 400 & 500 & $r_x$ ($\%$) \\
\hline
$N_{Ib/c}(\Delta t)$  &  8 & 9 & 9 & 13 & 13 & 15 & 17 & 18 & 18 & 20 & 21 & 26 & 30 & 42 & 68 & 81 & 96 & 12 \\
$N_{Ia}(\Delta t)$    & 10 & 23 & 30 & 42 & 51 & 64 & 77 & 85 & 98 & 108 & 118 & 131 & 164 & 213 & 338 & 440 & 519 & 66 \\
$N_{IIp}(\Delta t)$   & 2 & 4 & 8 & 14 & 16 & 19 & 19 & 21 & 22 & 26 & 27 & 30 & 39 & 54 & 82 & 103 & 124 & 16 \\
$N_{IIn}(\Delta t)$   & 1 & 2 & 4 & 6 & 8 & 9 & 9 & 9 & 10 & 11 & 11 & 11 & 14 & 21 & 30 & 38 & 51 & 6 \\
\hline
$N_{tot}(\Delta t)$   & 31 & 67 & 98 & 136 & 166 & 209 & 240 & 260 & 288 & 314 & 338 & 378 & 471 & 627 & 893 & 1139 & 1399 & 100\\
\hline
\hline
\end{tabular}}\label{tab:no3}
\end{table*}

\begin{figure*}
\centering
\includegraphics[scale=0.65,clip]{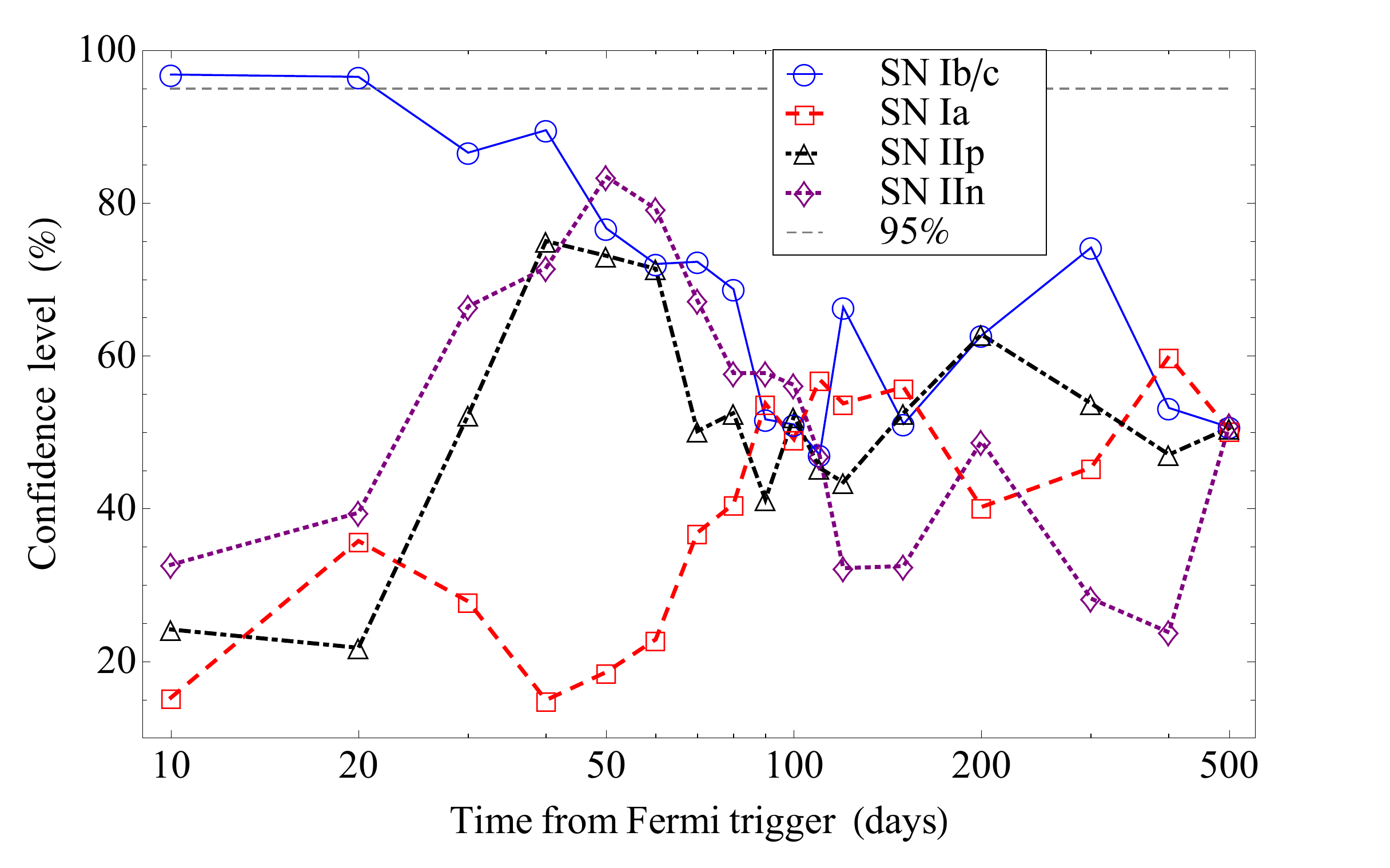}
\caption{The statistical significance of the GRB-SN occurrence as a function of the temporal window. This plot shows the significance of the deviation of SNe Ib/c in the time interval ($T_0 , T_0 + 20$ days) from the expected number of events assuming the relative proportion seen in the total SN sample.}
\label{fig:no2}
\end{figure*} 

If we assume a random distribution of SNe in the sky, the spatial GRB-SN association follows the Poisson statistic, $e^{-\lambda}\lambda^n/n!$, where $n$ is the number of observed associations and $\lambda$ is the expected number of positive events, in a chosen temporal window $\Delta t$. The expected number of positive events can be evaluated from $N_{tot}(\Delta t)$ (see last row in Table \ref{tab:no3}), times the percentage of each SN in the considered sample (see last column in Table \ref{tab:no3}). Therefore we have that $\lambda=N_{tot}(\Delta t)r_x$, where $x=\{$Ib/c,Ia,IIp,IIn$\}$. We have then compared it with the observations $N_x(\Delta t)$, and evaluated the corresponding confidence levels. The results of the computation are shown in Fig. \ref{fig:no2}. A simple comparison of significance tracks reported in Fig. 1 between SNe-Ibc and other SN types shows that, as expected, only SNe Ib/c within $\sim 30-40$ days after the GRB triggers are suggestive of the existence of physical associations with GRBs. From a simple application of Poissonian statistic in regime of small numbers \citep{Gehrels1986},  we derive a threshold of $\geq 95 \%$ confidence level, which corresponds to $\Delta t = 20$ days. In the following we will conservatively consider only associations between GRBs and SNe within $20$ days from the GRB trigger.

\section{The sample of GRBs-SNe Ib/c}

The list of GRB-SN Ib/c associations that our code has pinpointed is reported in Table \ref{tab:no1}, together with observational properties of the bursts and possibly related SNe. We found 5 cases. One of them, GRB 130702A - SN 2013dx is already known \citep{Singer2013}\footnote{If we relax the $z\leq0.2$ constraint, our code detect three more (well known) GRB-SN associations, 
specifically GRB 091127 - SN 2009nz \citep{Troja2009}, GRB 101219B - SN 2010ma \citep{Sparre2011}, GRB 130427A - SN 2013cq \citep{Xu2013,Melandri2014}}. For all SNe the redshift is determined from spectral observations of the host galaxy.    

\begin{table*}
\centering
\tiny{
\caption{Main parameters of the \textit{Fermi} GRB sample presented in this work and of the supernovae associated with these bursts. We also report the already known GRB-SN connection that we have found with our code in the last row of the table. * Nominal maximum value for the error radius of bursts detected by a single GBM detector.}
\label{tab:no1}
\begin{tabular}{l c c c c c c l c c c c}     
\hline
GRB & RA & DEC & Error & $T_{90}$ & Fluence & Peak flux & SN & date & RA & DEC & z  \\
& GBM & GBM & radius & & (0.01 - 1) MeV & (0.01 - 1) MeV &  & discovery & SN & SN &  \\
 & (deg) & (deg) & (deg) & (s) & (erg cm$^-2$) &  (photon cm$^-2$) & & & (deg) & (deg) &  \\ 
\hline
090320B & 183.4 & 49.8 & 9.5 & 29.2 & 1.67 $\times$ 10$^{-6}$& 4.35 $\pm$ 0.25 & 2009di & 2009 03 21 & 174.2411 & 45.0141 & 0.13 \\
090426B & 17.6 & -19.2 & 18.1 & 16.1 & 6.77 $\times$ 10$^{-7}$& 2.03 $\pm$ 0.18 & 2009em & 2009 05 05 & 8.6855 & -8.3993 & 0.006 \\
110911A & 258.58 & -66.98 & 50.0* & 8.96 & 5.94 $\times$ 10$^{-7}$& 2.38 $\pm$ 0.41 & 2011gw & 2011 09 15 & 112.0709 & -62.3552 & 0.01 \\
120121B & 235.67 & -39.34 & 7.9 & 18.4 & 1.95 $\times$ 10$^{-6}$ & 2.66 $\pm$ 0.21 & 2012ba & 2012 01 21 & 230.6047 & -38.2012 & 0.017 \\
\hline
130702A & 228.15 & 16.58 & 13.02 & 59 & 6.3 $\times$ 10$^{-6}$ & 7.03 $\pm$ 0.86 & 2013dx & 2013 07 08 & 217.3116 & 15.7740 & 0.145 \\
\hline
\end{tabular}}
\end{table*}

The values of $E_{iso}$ reported in  in Table \ref{tab:no2}  are derived from the spectral analysis of \textit{Fermi} GBM data of GRBs, using a Band function \citep{Band1993} as spectral model (see also Amati et al. 2008). We have considered Time-Tagged Events (TTE) \textit{Fermi} GBM spectra  which combine a high time resolution (up to 2 $\mu$s) with a good resolution in the spectral range. We fitted these spectra with the RMfit package\footnote{http://Fermi.gsfc.nasa.gov/ssc/data/analysis/rmfit/vc$\_$rmfit$\_$tutorial.pdf}. The value of $E_{iso}$ in the last column of the table shows that all events are low luminosity GRBs, unlike from so called ``cosmological'' GRBs, characterized by $E_{iso} \sim 10^{51}$ -- $10^{54}$ erg.  

\subsection{GRB 090320B - SN 2009di}

GRB 090320B was detected by the 10 and 11 \textit{Fermi} GBM detectors and also by Konus-WIND. The $T_{90}$ duration reported by \textit{Fermi} is 29.2 s, while unfortunately we do not have further information from Konus-WIND for this trigger. The possibly associated SN is SN 2009di, which was discovered on 21 March 2009, just one day after the detection by \textit{Fermi}, by the CRTS \citep{CBET1766}. At the moment of the discovery, the unfiltered magnitude of the SN was 18.6. Spectroscopy made with the 5.1m Palomar Hale telescope identified SN 2009di as a type Ic SN. The redshift of the SN was reported to be $z = 0.13$. The distance of the SN position and the \textit{Fermi} one is 7.8 degree, while the \textit{Fermi} error radius is about 9.5 degree.

\subsection{GRB 090426B - SN 2009em}

GRB 090426B was observed by the detectors 3 and 5 of \textit{Fermi} GBM, with a $T_{90}$ duration of 16.1 s. SN 2009em, associated with this GRB, was discovered by \citet{CBET1798} on 5 May 2009. Follow-up observations made 6 days late confirm the presence of an unfiltered magnitude 16.6 supernova. Further spectroscopic observations \citep{CBET1806,CBET1807} made around May 19 confirm the Ic nature of this SN, which corresponds to several known SNe Ic observed about one month from the maximum light, which plays against an association with GRB 090426B. The distance from the \textit{Fermi} position is 13.8 degree, to be compared with an error radius of 18 degree. 
The redshift of this source was measured to be $z = 0.006$, which corresponds to a comoving distance of 25.31 Mpc. 

\subsection{GRB 110911A - SN 2011gw}

This GRB triggered the \textit{Fermi} detectors number 2 and 10. However, the signal from detector number 2 was dominated by noise, so we have considered only the flux detected by number 10. This GRB was characterized by $T_{90} = 8.96$ s. 
SN 2011gw was discovered on 15 September by different observers, as an object of magnitude approximately 17.4 \citep{CBET2869}. A spectrum obtained one month later, on 20 October, at NTT telescope revealed the Ib/c nature of this supernova, and a cross-check with the GELATO library found a match with other SNe at about two months post maximum. The redshift of this SN was reported to be 0.01 while the distance between the center of \textit{Fermi} GBM detectors and the SN was of  48 degree, with an error radius of 50 degree. This large error box is due to the combination of two detectors that are located on the opposite sides of the \textit{Fermi} spacecraft and increases the probability of a casual association for this GRB-SN event.

\subsection{GRB 120121B - SN 2012ba}

GRB 120121B was detected by the \textit{Fermi} detectors number 3 and 5. The $T_{90}$ duration was of $18.4$ s. The best fit of the integrated spectrum of the GRB is a Band function with an intrinsic peak energy of $E_{p,i} = (92.2 \pm 12.2)$ keV. The SN associated to this GRB may be SN 2012ba. It was discovered on 21 January, the same day of the GRB trigger,  as an object of unfiltered magnitude 16.6 \citep{Pignata2012}, still in rising phase. A spectrum obtained on 2 March (40 days after the discovery) with the 6.5-m Magellan II Clay telescope and then cross-correlated with the SNID libraries of SN spectra, showed a match with a type Ic SN more than 15 days after maximum. The redshift of the SN, $z = 0.017$ associated with the observed peak magnitude of 15.9, eleven days after the SN discovery \citep{Pignata2012}, implied an absolute magnitude at maximum of -18.5, which is an upper limit to the intrinsic luminosity, considering the correction for dust extinction. This result suggests that SN 2012ba is a very luminous SN Ic, with an absolute magnitude similar to that of SN 2010bh, $R_{abs} \approx -18.5$ \citep{Bufano2012} or even brighter, similarly to SN 1998bw
$R_{abs} \approx -19$ \citep{Patat2001}. The distance between the SN position and the \textit{Fermi} center was of 4.1 degree, inside the \textit{Fermi} error radius of 7.9 degree.

\begin{table}
\centering
\tiny{
\caption{Results of the spectral fits of \textit{Fermi} GBM observations for the 4 GRBs with evidence of association with a SN Ic.}
\label{tab:no2}
\begin{tabular}{l c c c c}     
\hline
GRB & $\alpha$ & $\beta$ & $E_{peak}$ & $E_{iso}$ \\
&  &  & $(keV)$  & $(erg)$  \\ 
\hline
090320B & -0.65 $\pm$ 0.35 & -2.42 $\pm$ 0.30 & 62.6 $\pm$ 12.0 & 9.13 $\times$ 10$^{49}$\\
090426B & -0.50 $\pm$ 3.12 & -1.65 $\pm$ 0.15 & 39.9 $\pm$ 76.9 & 1.94 $\times$ 10$^{47}$\\
110911A & -0.47 $\pm$ 0.50 & -1.36 $\pm$ 0.18 & 44.8 $\pm$ 20.1 & 6.22 $\times$ 10$^{47}$\\
120121B & -0.73 $\pm$ 0.21 & -2.95 $\pm$ 0.89 & 92.2 $\pm$ 12.2 & 1.39 $\times$ 10$^{48}$\\
\hline
\end{tabular}}
\end{table}

\section{Discussions}

Our analysis discovered 5 GRB-SN coincidences within $z\leq0.2$, and one of them was already known to be a physical 
association between GRB and SN (GRB 130702A-SN 2013dx \citep{Singer2013}). We note that the afterglow of GRB 130702A has been found by the authors of the above cited work upon searching 71 deg$^2$ surrounding the \textit{Fermi}-GBM localization. This result further strengthen the reliability of the adopted methodology.

After discussion of the data, we found that SN 2012ba is the only ``bona fide'' candidate for being physical associated with a GRB (120121B). SN 2012ba was of type Ic and reached quickly a very bright maximum magnitude $R_{abs} \simeq -19$, about 11 days after the GRB trigger \citep{CBET3025}, which is very similar to the typical rising time of SNe associated with GRBs \citep{Bufano2012}. To date there are only two other SNe associated with GRBs and classified as ``Ic'' (rather than ``broad lines'' Ic or Hypernovae): SN 2002lt, associated to GRB 021211 \citep{dellavalle2003}, and SN 2013ez, associated to GRB 130215A \citep{Cano2014}. However, these observations do not imply that GRBs may be associated with ``standard type-Ic SNe''. We note that in all three cases, 2012ba, 2002lt and 2013ez, SN spectra were secured 20-40 days past maximum, therefore even if the pre-maximum spectra showed significantly broader lines, than observed in the post-maximum spectra, this difference shortly vanished after maximum (if the SN ejecta carry little mass) such that it is not easy to distinguish between the two types of SNe. The isotropic energy of this \textit{Fermi} GRB-SN candidate is $E_{iso}$ = 1.39 $\times$ 10$^{48}$ erg, which implies that this burst belongs to the low-luminosity subclass of GRBs \citep{Guetta2007,PiranBromberg2013,Tsutsui}}. Now, we are in the position to independently estimate, admittedly on the very scanty statistic of one single object, the rate $\rho_0$ of local low-energetic long GRBs-- type Ic SNe. Following \citet{Soderberg2006Nature} and \citet{Guetta2007}, we have computed the photon peak flux $f_p$ in the energy band $1$--$1000$ keV and the corresponding threshold peak flux, following the analysis of \citet{Band2003} for GRB 120121B. In this way we have evaluated the maximum redshift $z_{max}$ at which this burst would have detected, $z = 0.0206$, and then the corresponding maximum comoving volume $V_{max}$.

The empirical rate can then be written as 
\begin{equation}
\label{rate1}
\rho_0=\frac{N_{LE}}{V_{max} f_F T}\ ,
\end{equation}
where $N_{LE} = 1$ is the number of found physical connections, $f_F \approx 0.76$ the average ratio of \textit{Fermi} solid angle over the total one, and $T=6$ years the \textit{Fermi} observational period.  We infer a local rate for this GRB--SN Ic events of $\rho_0=77^{+289}_{-73}$Gpc$^{-3}$yr$^{-1}$, where the errors are determined from the $95\%$ confidence level of the Poisson statistic \citep{Gehrels1986}. There is growing body of evidence that low luminosity GRBs are less beamed that high luminosity GRBs, indeed  $f_b^{-1}$ is of the order of 10,  or less \citep[see e.g.][]{Guetta2007}. After taking into account this correction we derive $\rho_{0,b}\leq 770$Gpc$^{-3}$yr$^{-1}$, which is consistent with $\rho_0=380^{+620}_{-225}$Gpc$^{-3}$yr$^{-1}$ in \citet{Guetta2007}, $325^{+352}_{-177}$Gpc$^{-3}$yr$^{-1}$ in \citet{Liang2007}, and 
 $230^{+490}_{-190}$Gpc$^{-3}$yr$^{-1}$ in \citet{Soderberg2006Nature}.

This analysis confirms the existence of a class of more frequent low-energetic GRBs--SNe Ic, whose rate is larger than the one obtained extrapolating at low redshifts the rate for high-energetic bursts, i.e., $\rho=1.3^{+0.7}_{-0.6}$Gpc$^{-3}$yr$^{-1}$ \citep{Wanderman2010}.

\section{Conclusions}

This paper presents the results of an analysis dedicated to find possible connections between long GRBs listed in the \textit{Fermi}-GBM catalog and SNe. Our analysis was motivated by the fact that we expected, on statistical basis, to find in the \textit{Fermi} catalog a minimum of $1$ up to $7$ one GRB-SN connections within $z < 0.2$.  From our analysis the following results emerge:
\begin{itemize}

\item we have found a total number of $5$ possible connections at $z\leq0.2$.  One of them was already known as physical GRB-SN associations. After discussing the remaining $4$ cases, we found that only GRB 120121B is very likely physically connected with SN
2012ba. This result of two observed GRBs-SNe is fully consistent with our initial estimate of $1$--$7$ low-$z$ events to be found in the \textit{Fermi} catalogue;

\item the very low redshift at which GRB 120121B/SN 2012b is observed implies a small isotropic energy emitted during the GRB, $E_{iso} = 1.39 \times 10^{48}$ erg. From this single connection, we compute the rate of \textit{Fermi} low-luminosity GRBs connected with SNe to be $\rho_0=77^{+289}_{-73}$Gpc$^{-3}$yr$^{-1}$. If we consider an additional correction, due to a beaming in the low luminosity GRB emission, $f_b^{-1}$ of the order of $10$ \citep{Guetta2007}, we obtain for the \textit{Fermi} rate $\rho_{0,b} \leq 770$Gpc$^{-3}$yr$^{-1}$, which is consistent with $\rho_0=380^{+620}_{-225}$Gpc$^{-3}$yr$^{-1}$ in \citet{Guetta2007}, $325^{+352}_{-177}$Gpc$^{-3}$yr$^{-1}$ in \citet{Liang2007}, and $230^{+490}_{-190}$Gpc$^{-3}$yr$^{-1}$ in \citet{Soderberg2006Nature};

\item if we consider a continuous time coverage, including previous analysis from Beppo-SAX (7 years, 1 connection -- GRB 980425, \citealp{Galama1998}) and \textit{Swift} (9 years, 2 connections -- GRB 060218, \citealp{Campana2006}, and GRB 100316D, \citealp{Bufano2012}), we obtain a comprehensive rate of $\rho^{tot}_0=31^{+40}_{-20}$Gpc$^{-3}$yr$^{-1}$, which becomes $\rho^{tot}_{0,b} = 310^{+400}_{-200}$Gpc$^{-3}$yr$^{-1}$, assuming $f_b^{-1}$ of the order of $10$.

\item On the basis of the annual rate of \textit{Fermi} GRBs ($238$ GRBs/year), and of the expected number of \textit{Fermi}-GBM bursts associated with low-z SNe ($1$--$7$ GRBs) in 6 years of observations, we estimate that in the next $4$ years \textit{Fermi}-GBM could detect $\sim 1$--$4$ GRBs-SNe within $z \leq 0.2$.
\end{itemize}



\begin{acknowledgements}

We are grateful to Remo Ruffini, which provided support for the final outcome of this work. MK, ME, GBP and LL are supported by the Erasmus Mundus Joint Doctorate Program by grant Nos.  2013-1471, 2012-1710, 2011-1640 and 2013-1471, respectively, from the EACEA of the European Commission.

\end{acknowledgements}

\bibliographystyle{aa}

\end{document}